\documentclass[12pt]{article}

\usepackage{amsmath,amssymb,cite}
\usepackage{graphicx}

\makeatletter
 
  \@addtoreset{equation}{section}
 \makeatother

\newcommand{\be}{\begin{equation}}
\newcommand{\ee}{\end{equation}}
\newcommand{\bea}{\begin{eqnarray}}
\newcommand{\eea}{\end{eqnarray}}
\newcommand{\beann}{\begin{eqnarray*}}
\newcommand{\eeann}{\end{eqnarray*}}

\newcommand{\ba}{\begin{array}}
\newcommand{\ea}{\end{array}}

\newcommand{\Tr}{\mathop{\rm Tr}}

\newcommand{\diag}{\mathop{\rm diag}\nolimits}

\newcommand{\e}{\epsilon}

\newcommand{\del}{\partial}

\newcommand{\calZ}{{\cal Z}}

\def\XXint#1#2#3{{\setbox0=\hbox{$#1{#2#3}{\int}$} 
\vcenter{\hbox{$#2#3$}}\kern-.5\wd0}}

\begin{document}

\setlength{\oddsidemargin}{0cm}
\setlength{\baselineskip}{7mm}

\begin{titlepage}
\renewcommand{\thefootnote}{\fnsymbol{footnote}}
\begin{normalsize}
\begin{flushright}
\begin{tabular}{l}
OU-HET 618\\
KEK-TH-1289\\
TU-830 \\
November 2008
\end{tabular}
\end{flushright}
  \end{normalsize}

~~\\

\vspace*{0cm}
    \begin{Large}
       \begin{center}
         {Two-Dimensional Gauge Theory and Matrix Model}
       \end{center}
    \end{Large}
\vspace{0.7cm}

\begin{center}
Goro I{\sc shiki}$^{1),2)}$\footnote
            {
e-mail address : 
ishiki@post.kek.jp},
Kazutoshi O{\sc hta}$^{3)}$\footnote
            {
e-mail address : 
kohta@tuhep.phys.tohoku.ac.jp}, 
Shinji S{\sc himasaki}$^{1)}$\footnote
            {
e-mail address : 
shinji@het.phys.sci.osaka-u.ac.jp}
    {\sc and}
Asato T{\sc suchiya}$^{4)}$\footnote
           {
e-mail address : 
satsuch@ipc.shizuoka.ac.jp}\\
      
\vspace{0.7cm}
                    
       $^{1)}$ {\it Department of Physics, Graduate School of  
                     Science}\\
               {\it Osaka University, Toyonaka, Osaka 560-0043, Japan}\\
      \vspace{0.3cm}
      $^{2)}$ {\it Institute of Particle and Nuclear Studies}\\
               {\it High Energy Accelerator Research Organization (KEK)}\\
               {\it 1-1 Oho, Tsukuba, Ibaraki 305-0801, Japan}\\
      \vspace{0.3cm}
       $^{3)}$ {\it High Energy Theory Group, Department of Physics}\\
               {\it Tohoku University, Sendai 980-8578, Japan}\\
      \vspace{0.3cm}
      $^{4)}$ {\it Department of Physics, Shizuoka University}\\
               {\it 836 Ohya, Suruga-ku, Shizuoka 422-8529, Japan}
               
\end{center}

\vspace{0.7cm}

\begin{abstract}
\noindent
We study a matrix model obtained by dimensionally reducing Chern-Simon theory on $S^3$.
We find that the matrix integration is decomposed into sectors classified by the representation
of $SU(2)$. We show that the $N$-block sectors reproduce $SU(N)$ 
Yang-Mills theory on $S^2$ as the matrix size goes to infinity.
\end{abstract}
\vfill
\end{titlepage}
\vfil\eject

\setcounter{footnote}{0}

\section{Introduction}
Matrix models have been proposed as non-perturbative formulation of 
superstring or M-theory \cite{BFSS,IKKT,DVV}. Since low energy physics predicted by 
string theory depends on topological aspects of compactification,
it is relevant to investigate how they are incorporated in matrix models.
The topological field theories have been developed to efficiently
describe the topological aspects of field theories.
It is, therefore, worthwhile to study realization of 
the topological field theories in matrix models.

Hinted by the work \cite{Ishiki:2006yr},
the authors of \cite{Ishii:2007sy} found the following classical relationships among
Chern-Simons (CS) theory on $S^3$, two-dimensional Yang-Mills (2d YM) on $S^2$ and
a matrix model. The latter two theories are obtained by dimensionally reducing
the first theory. 
The theory around each multiple monopole background of 2d YM
is obtained by expanding the matrix model around a certain multiple fuzzy sphere background in the 
continuum limit (see also \cite{Ishii:2008tm}).
CS theory is obtained by applying an extension of
compactification in matrix models developed in \cite{Ishiki:2006yr,Ishii:2007ex,Ishii:2008tm} 
to the theory around a multiple monopole background
of 2d YM. Eventually,  CS theory is obtained by expanding the matrix model around a certain
multiple fuzzy sphere background and imposing the orbifolding condition.
2d YM is also viewed as BF theory with a mass term on $S^2$.
The matrix model takes the form of the superpotential for ${\cal N}=1^*$ theory.
The classical relationships between CS on $S^3$ and 2d YM on $S^2$ are generalized to
those between CS theory on a $U(1)$ bundle over a Riemann surface $\Sigma_g$ of genus $g$
and 2d YM on $\Sigma_g$.

In this Letter, we show that 2d YM on $S^2$ 
is obtained from the matrix model also at quantum level.
We find that the matrix integration is decomposed into sectors classified by
the representation of $SU(2)$. We show that the $N$-block sectors reproduce the partition function
of 2d $SU(N)$ YM on $S^2$. 

It has been already shown in \cite{Steinacker:2003sd,Steinacker:2007iq}
that different types of matrix models give 2d YM on $S^2$. 
Moreover, the authors of \cite{Steinacker:2007iq}
have shown that 
the localization works also for the matrix model in the same way as it works for the continuum 2d YM.
We hope to elucidate the relation of our work with \cite{Steinacker:2003sd,Steinacker:2007iq} in the future.

This Letter is organized as follows. 
In section 2, we briefly review part of the results in \cite{Ishii:2007sy}, which are associated with
the present work.
In section 3, we reduce the path-integral in the matrix model to the integral over the eigenvalues of
a single matrix, which is decomposed into the sectors classified by the  representation of $SU(2)$.
In section 4, we show that part of the above sectors reproduce 2d YM on $S^2$.
Section 5 is devoted to conclusion and outlook. In appendix, we summarize some useful properties
of $S^3$ and $S^2$.

\section{Classical relationships among CS theory, 2d YM and a matrix model}
In this section, we briefly review only part of the results in \cite{Ishii:2007sy} which are concerned 
with the present Letter.
We start with CS theory on $S^3$ with the gauge group $U(M)$:
\begin{align}
S_{CS}=\frac{k}{4\pi}\int_{S^3}\mbox{Tr}\left(A\wedge dA+\frac{2}{3}A\wedge A\wedge A\right).
\label{CS on S^3}
\end{align}
We expand the gauge field in terms of the right-invariant 1-form defined in (\ref{right-invariant 1-form}) as
\begin{align}
A=iX_iE^i.
\end{align}
Then, we rewrite (\ref{CS on S^3}) as
\begin{align}
S_{CS}=-\frac{k}{4\pi}\int \frac{d\Omega_3}{(\mu/2)^3}
\mbox{Tr}\left(i\mu\epsilon^{ijk}X_i{\cal L}_jX_k+\mu X_i^2+\frac{2i}{3}\epsilon^{ijk}X_iX_jX_k\right),
\end{align}
where ${\cal L}_i$ is the Killing vector dual to $E^i$ and defined in (\ref{definition of Killing vector}).

By dropping the derivative of the fiber direction $y$, we obtain a gauge theory on $S^2$:
\begin{align}
S_{BF}=-\frac{1}{g_{BF}^2\mu}\int \frac{d\Omega_2}{\mu^2}
\mbox{Tr}\left(i\mu\epsilon^{ijk}X_iL^{(0)}_jX_k+\mu X_i^2+\frac{2i}{3}\epsilon^{ijk}X_iX_jX_k\right),
\label{BF with mass term on S^2}
\end{align}
where $g_{BF}^2=1/k$\footnote{While $k$ in (\ref{CS on S^3}) must be integer, such a restriction is not
imposed on $k$ in (\ref{BF with mass term on S^2}).} and $L^{(0)}_i$ are the angular momentum operators
on $S^2$ given in (\ref{monopole angular momentum}) with $q=0$.
In order to see that (\ref{BF with mass term on S^2}) is BF theory with a mass term, we 
define $L^{(0)\mu}_i\;\;(\mu=\theta, \varphi)$ by $L^{(0)}_i=L^{(0)\mu}_i\partial_{\mu}$ and introduce
$N_i\;\;(i=1,2,3)$ given by
\begin{align}
N_1=\sin\theta\cos\varphi, \;\;\; N_2=\sin\theta\sin\varphi,\;\;\; N_3=\cos\theta.
\end{align}
Then, it is easy to see that $L^{(0)\mu}_i$ and $N_i$ satisfy the following relations:
\begin{align}
&L^{(0)\mu}_iL^{(0)\nu}_i=-g^{\mu\nu}, \;\;\;
N_iN_i=1, \;\;\;
L^{(0)\mu}_iN_i=0, \nonumber\\
&L^{(0)\mu}_i\partial_{\mu}L^{(0)\nu}_j-L^{(0)\mu}_j\partial_{\mu}L^{(0)\nu}_i=i\epsilon_{ijk}L^{(0)\nu}_k,
\nonumber\\
&L^{(0)\mu}_i\partial_{\mu}N_j-L^{(0)\mu}_j\partial_{\mu}N_i=2i\epsilon_{ijk}N_k, \nonumber\\
&\epsilon_{ijk}N_iL^{(0)\mu}_jL^{(0)\nu}_k=-\epsilon^{\mu\nu},
\label{relations}
\end{align}
where $g^{\mu\nu}$ and $\epsilon^{\mu\nu}$ can be read off from (\ref{metric of S^2}).
We expand $X_i$ as \cite{Kitazawa:2002xj,Ishii:2008tm}
\begin{align}
X_i=\mu (i L^{(0)\mu}_ia_{\mu}+N_i\chi).
\end{align}
$a_{\mu}$ and $\chi$ turn out to be the gauge field and a scalar filed on $S^2$, respectively.
By using the relations (\ref{relations}), we can show that (\ref{BF with mass term on S^2}) is equivalent to
\begin{align}
S_{BF}=-\frac{\mu^2}{g_{BF}^2}\int\frac{d\Omega_2}{\mu^2}
\mbox{Tr}\left(\chi\epsilon^{\mu\nu} f_{\mu\nu}-\chi^2\right),
\label{BF with mass term on S^2 2}
\end{align}
where $f_{\mu\nu}=\del_\mu a_\nu-\del_\nu a_\mu+i[a_\mu,a_\nu]$ is the field strength.
Indeed, the first term is the BF term and the second term is a mass term.

By integrating $\chi$ out in (\ref{BF with mass term on S^2 2}), we obtain
2d YM on $S^2$:
\begin{align}
S_{2dYM}=\frac{\mu^4}{g_{YM}^2}\int \frac{d\Omega_2}{\mu^2}
\mbox{Tr}\left(\frac{1}{4}f^{\mu\nu}f_{\mu\nu}\right),
\label{2d YM}
\end{align}
where $1/g_{YM}^2=-2/(g_{BF}^2\mu^2)$.

By dropping all the derivatives in (\ref{BF with mass term on S^2}) and rescale $X_i$ as 
$X_i\rightarrow \mu X_i$, we obtain ${\cal N}=1^{\ast}$ matrix model:
\begin{align}
S=-\frac{1}{g^2}\mbox{Tr}\left(X_i^2+\frac{i}{3}\epsilon^{ijk}X_i[X_j,X_k]\right),
\label{N=1^* matrix model}
\end{align}
where $1/g^2=4\pi/g_{BF}^2$.
In the sense of the Dijkgraaf-Vafa theory \cite{Dijkgraaf:2002fc}, this matrix model is regarded as
a mass deformed superpotential of ${\cal N}=4$ supersymmetric Yang-Mills
theory (super-YM), which gives the so-called ${\cal N}=1^*$ theory. 
We call the matrix model (\ref{N=1^* matrix model}) the ${\cal N}=1^*$ matrix model in this Letter.

Inversely, we can obtain the BF theory with the mass term from the matrix model as follows.
The matrix model (\ref{N=1^* matrix model}) possesses the following classical solution,
\begin{align}
\hat{X}_i=L_i=\left(
\begin{array}{cccc}
L^{[j_1]}_i &&& \\
& L^{[j_2]}_i && \\
&& \ddots & \\
&&& L^{[j_N]}_i
\end{array} \right),
\label{matrix background}
\end{align}
where $L^{[j_s]}_i \;\;(s=1,\cdots,N)$ are the spin $j_s$ representation of the $SU(2)$ generators 
obeying $[L^{[j_s]}_i,L^{[j_s]}_j]=i\epsilon_{ijk}L^{[j_s]}_k$, and the relation $\sum_{s=1}^N(2j_s+1)=M$ is 
satisfied. We label the blocks by $s$.
We put $2j_s+1=N_0+n_s$ with $N_0$ and $n_s$ integers and take the limit in which 
\begin{align}
N_0\rightarrow\infty \;\;\; \mbox{with} \;\;\; 
\frac{N_0}{g^2}=\frac{4\pi}{g_{BF}^2}=-\frac{8\pi^2}{g_{YM}^2A}=\mbox{fixed},
\label{limit}
\end{align}
where $A=4\pi/\mu^2$ is the area of $S^2$.
Then,  we can show classically \cite{Ishii:2007sy} that
the theory around (\ref{matrix background}) is equivalent to the theory around
the following classical solution of (\ref{BF with mass term on S^2}),
\begin{align}
\mu L^{(0)}_i+\hat{X}_i=\mu\mbox{diag}(L^{(q_1)}_i,L^{(q_2)}_i,\cdots,L^{(q_N)}_i),
\end{align}
where $q_s=n_s/2$, and $L^{(q_s)}_i$ are the angular momentum operators in the presence of a monopole
with the monopole charge $q_s$, which are given in (\ref{monopole angular momentum}). 
This theory can also be viewed as the theory around the following classical solution
of (\ref{BF with mass term on S^2 2}),
\begin{align}
&\hat{\chi}=-\mbox{diag}(q_1,q_2,\cdots,q_N), \nonumber\\
&\hat{a}_{\theta}=0, \nonumber\\
&\hat{a}_{\varphi}=(\cos\theta \mp 1) \hat{\chi},
\label{monopole solution}
\end{align}
where the upper sign is taken in the region $0\leq \theta < \pi$ and the lower sign
in the region $0 < \theta \leq \pi$, and $\hat{a}_{\theta}$ and $\hat{a}_{\varphi}$ represent
the monopole configuration.

\section{Exact integration of the partition function}
In this section, we evaluate the partition function of (\ref{N=1^* matrix model}). We reduce
the path-integral in the matrix model  to the integral over the eigenvalues of a single matrix.
In (\ref{N=1^* matrix model}), we redefine the matrices as 
\begin{align}
Z=X_1+iX_2, \;\;\; Z^{\dagger}=X_1-iX_2,\;\;\; \Phi=X_3.
\label{redefinition of matrices}
\end{align}
$Z$ is an $M\times M$ complex matrix while $\Phi$ is an $M\times M$ hermitian matrix.
Using (\ref{N=1^* matrix model}) and (\ref{redefinition of matrices}), 
we define the partition function of ${\cal N}=1^{\ast}$ matrix model (\ref{N=1^* matrix model}) by
\begin{align}
{\cal Z}=\lim_{\e\to0}\int d\Phi dZ dZ^{\dagger} e^{-\frac{i}{g^2}\Tr(Z[\Phi,Z^{\dagger}]+(1-i\epsilon)ZZ^{\dagger}+\Phi^2)},
\label{partition function of N=1^* matrix model}
\end{align}
where we introduce the `$-i\epsilon$' term in the action to make the integral converge.
Integral over $Z$ and $Z^{\dagger}$ leads to a one matrix model with respect to $\Phi$
\cite{Kazakov:1998ji,Dijkgraaf:2002fc,Dorey:2001qj}
\[
{\cal Z} = \lim_{\e\to0}\int d\Phi \frac{1}{\det([\Phi,\cdot]+1-i\e)}e^{-\frac{i}{g^2}\Phi^2},
\]
where $[\Phi,\cdot]$ represents an adjoint action. Furthermore, if we diagonalize $\Phi$ as $\Phi=\mbox{diag}(\phi_1,\phi_2,\cdots,\phi_M)$, the matrix integral reduces integrals over the eigenvalues $\phi_i$
\begin{align}
{\cal Z}=\frac{1}{M !}\lim_{\e\to0}\int \prod_i d\phi_i \prod_{i\neq j}\frac{\phi_i-\phi_j}{\phi_i-\phi_j+1-i\epsilon} 
e^{-\frac{i}{g^2}\sum_i\phi_i^2}.
\label{partition function of N=1^* matrix model 2}
\end{align}
where 
$\prod_{i\neq j}(\phi_i-\phi_j)$ in the numerator of the integrand comes from the Vandermonde determinant
owing to the diagonalization of $\Phi$.

As a simple example, we consider the $M=2$ case.
In this case, (\ref{partition function of N=1^* matrix model 2}) is explicitly written as
\begin{align}
{\cal Z}=\frac{1}{2}\lim_{\e\to 0}\int d\phi_1 d\phi_2 
\frac{(\phi_1-\phi_2)(\phi_2-\phi_1)}{(\phi_1-\phi_2+1-i\epsilon)(\phi_2-\phi_1+1-i\epsilon)}
e^{-\frac{i}{g^2}(\phi_1^2+\phi_2^2)}.
\label{2 times 2}
\end{align}
In what follows, we frequently use the identity
\begin{align}
\lim_{\e\to 0}\frac{1}{x-i\epsilon}=P.V.\:\frac{1}{x}+i\pi\delta(x),
\label{principal value}
\end{align}
where $P.V.$ stands for Cauchy's principal value of an integral.
Applying (\ref{principal value}) to (\ref{2 times 2}) leads to
\begin{align}
{\cal Z}&=\frac{1}{2}P.V.\!\int d\phi_1 d\phi_2\frac{(\phi_1-\phi_2)(\phi_2-\phi_1)}{(\phi_1-\phi_2+1)(\phi_2-\phi_1+1)}
e^{-\frac{i}{g^2}(\phi_1^2+\phi_2^2)} \nonumber\\
&\quad -\frac{i\pi}{4}\int d\phi_1 \left(
e^{-\frac{i}{g^2}(\phi_1^2+(\phi_1+1)^2)}+e^{-\frac{i}{g^2}(\phi_1^2+(\phi_1-1)^2)}
\right) \nonumber\\
&=\frac{1}{2}P.V.\!\int d\phi_1 d\phi_2\frac{(\phi_1-\phi_2)(\phi_2-\phi_1)}{(\phi_1-\phi_2+1)(\phi_2-\phi_1+1)}
e^{-\frac{i}{g^2}(\phi_1^2+\phi_2^2)} \nonumber\\
&\quad -\frac{i\pi}{2} e^{-\frac{i}{2g^2}}\int d\phi 
e^{-\frac{2i}{g^2}\phi^2}.
\end{align}

We generalize the above calculation to the case of arbitrary $M$.
We apply (\ref{principal value}) to the factor in the integrand of 
(\ref{partition function of N=1^* matrix model 2}),
\begin{align}
\prod_{i\neq j}\frac{\phi_i-\phi_j}{\phi_i-\phi_j+1-i\epsilon},
\end{align}
and obtain the sum of the terms, each of which includes some delta functions.
It is easily seen that any term giving non-vanishing contribution must be proportional to
\begin{align}
&\left(\frac{-i\pi}{2}\right)^{\sum_{s=1}^N2j_s}\times
\delta(\phi_1^{(1)}-\phi_2^{(1)}-1)\delta(\phi_2^{(1)}-\phi_3^{(1)}-1)\cdots \delta(\phi_{2j_1}^{(1)}
-\phi_{2j_1+1}^{(1)}-1)
\nonumber\\
&\times 
\delta(\phi_1^{(2)}-\phi_2^{(2)}-1)\delta(\phi_2^{(2)}-\phi_3^{(2)}-1)\cdots \delta(\phi_{2j_2}^{(2)}
-\phi_{2j_2+1}^{(2)}-1)
\nonumber\\
&\times \cdots \nonumber\\
&\times 
\delta(\phi_1^{(N)}-\phi_2^{(N)}-1)\delta(\phi_2^{(N)}-\phi_3^{(N)}-1)\cdots 
\delta(\phi_{2j_N}^{(N)}-\phi_{2j_N+1}^{(N)}-1),
\label{product of delta functions}
\end{align}
where we have reordered and relabeled 
the eigenvalues of $\Phi$, $\phi_i \;(i=1,\cdots,M))$, as 
\begin{align}
\Phi=\diag(\phi_1^{(1)},\cdots,\phi_{2j_1+1}^{(1)},
\phi_1^{(2)},\cdots,\phi_{2j_2+1}^{(2)},\cdots,
\phi_1^{(N)},\cdots,\phi_{2j_N+1}^{(N)}),
\label{Phi}
\end{align}
with $\sum_{s=1}^N(2j_s+1)=M$, such that the form of (\ref{product of delta functions}) is obtained.
$\phi^{(s)}_i\;(i=1,\cdots,2j_s+1)$ represents the $i$-th component of the $s$-th block. 
(\ref{product of delta functions}) and (\ref{Phi}) specify an 
$M$-dimensional irreducible representation of $SU(2)$ consisting of $N$ blocks as seen in (\ref{matrix background}),  
with a $U(1)$ degree of freedom in each block. We label the irreducible representation by $r$ and
denote the $U(1)$ part in the $s$-th block by $a_s$, putting $a_s\equiv\phi_{2j_s+1}^{(s)}+j_s$. 
Then, we find that the contribution of (\ref{product of delta functions}) 
to (\ref{partition function of N=1^* matrix model}) is 
\begin{align}
&{\cal N}_r(-i\pi)^{M-N}\prod_{s=1}^N\frac{1}{2j_s+1}
 P.V.\! \int \prod_{s=1}^Nda_s
\prod_{s\neq t}\prod_{m_s=-j_s}^{j_s}\prod_{m_t=-j_t}^{j_t}
\frac{a_s+m_s-a_t-m_t}{a_s+m_s-a_t-m_t+1} \nonumber\\
&\qquad\qquad\qquad\qquad\qquad \times
e^{-\frac{i}{g^2}\sum_{s=1}^N\sum_{m_s=-j_s}^{j_s}(a_s+m_s)^2},
\label{part of Z}
\end{align}
where 
\begin{align}
{\cal N}_r=\prod\frac{1}{(\sharp\mbox{ of blocks with the same length})!}.
\end{align}
and the other factor in (\ref{part of Z}) is obtained from the following calculation:
\begin{align}
\prod_{s=1}^N\left(\frac{-i\pi}{2}\right)^{2j_s}
\sum_{s=1}^N\prod_{k=2}^{2j_s}\left(\frac{k^2}{k^2-1}\right)^{2j_s-k+1}
=(-i\pi)^{M-N}\prod_{s=1}^N\frac{1}{2j_s+1}.
\end{align}

We further do some algebra for the exponent in (\ref{part of Z}):
\begin{align}
\sum_{s=1}^N\sum_{m_s=-j_s}^{j_s}(a_s+m_s)^2
=\sum_{s=1}^N\left((2j_s+1)a_s^2+\frac{1}{3}j_s(j_s+1)(2j_s+1)\right).
\end{align}
By composing the angular momenta, we also evaluate the product appearing in (\ref{part of Z}):
\begin{align}
\prod_{s\neq t}\prod_{m_s=-j_s}^{j_s}\prod_{m_t=-j_t}^{j_t}
\frac{a_s+m_s-a_t-m_t}{a_s+m_s-a_t-m_t+1}
&=\prod_{s\neq t}\prod_{J=|j_s-j_t|}^{j_s+j_t}\prod_{m=-J}^J 
\frac{m+a_s-a_t}{1+m+a_s-a_t}     \nonumber\\
&=\prod_{s\neq t}\prod_{J=|j_s-j_t|}^{j_s+j_t}\frac{J+a_s-a_t}{-J-1+a_s-a_t} \nonumber\\
&=\prod_{s< t}\frac{(j_s-j_t)^2-(a_s-a_t)^2}{(j_s+j_t+1)^2-(a_s-a_t)^2}.
\end{align}
Gathering all the above results, we eventually find that (\ref{partition function of N=1^* matrix model 2}) results in
\begin{align}
{\cal Z}=&\sum_{r} {\cal N}_r (-i\pi)^{M-N}\prod_{s=1}^N\frac{1}{2j_s+1}
\;e^{-\frac{i}{3g^2}\sum_{s=1}^N\mbox{tr}(L_i^{[j_s]})^2} \nonumber\\
&\times P.V. \! \int \prod_{s=1}^Nda_s 
\prod_{s< t}\frac{(j_s-j_t)^2-(a_s-a_t)^2}{(j_s+j_t+1)^2-(a_s-a_t)^2}
e^{-\frac{i}{g^2}\sum_{s=1}^N(2j_s+1)a_s^2},
\label{partition function of N=1^* matrix model 3}
\end{align}
where $L^{[j_s]}_i$ is the spin $j_s$ representation of the $SU(2)$ generators seen in
(\ref{matrix background}). Thus the partition function of the ${\cal N}=1^*$ matrix model is 
decomposed into the sectors classified by the irreducible representation of $SU(2)$.
Indeed, it is ensured by $P.V.$ that the whole integral region of $a_s$ are decomposed into these sectors
without overlap, which means that the full matrix integral over $X_1,\;X_2,\;X_3$ is decomposed into these sectors without overlap.

\section{Relation to Continuum Field Theory}

In this section, we reproduce 2d YM on $S^2$ from the
${\cal N}=1^*$ matrix model in the large matrix size limit.
As we will see, the number of the matrix blocks in the irreducible representation of $SU(2)$, $N$, 
corresponds to the rank of the gauge group of 2d YM.
Since there is no overlap between the decomposed sectors in the matrix model partition function 
(\ref{partition function of N=1^* matrix model 3}),
we can extract the sectors with a fixed $N$.
But one question arises: What type of the partition of blocks is dominated in the large matrix size limit with fixed $N$?

To see this, let us investigate the ``potential'' in the $N$-block sectors in 
the partition function (\ref{partition function of N=1^* matrix model 3})
\[
V(\vec{a},\vec{d},\lambda)=\sum_{s=1}^{N} \left(d_s a_s^2 +\frac{1}{12}d_s(d_s^2-1)\right)+\lambda(\sum_{s=1}^N d_s - M),
\]
where we put $d_s=2j_s+1$ and 
$\lambda$ is a Lagrange multiplier for the constraint $\sum_{s=1}^{N} d_s =M$.
This potential is minimized at $a_s=0$ and $d_s=M/N$ for ${}^\forall s$, that is, a configuration of almost equal size blocks
is dominated.

Thus we now consider the fluctuation around the dominated configuration
\[
d_s \equiv N_0 + n_s,
\]
where $M=NN_0$ and $\sum_{s=1}^Nn_s=0$.
In the large matrix size limit, we take
the limit (\ref{limit}) with fixed $N$, which reduces 
the $N$-block sectors to
\be
\calZ_{N} =
C\sum_{\sum_sn_s=0}
\int_{\sum_sa'_s=0} \prod_{s=1}^N da'_s
\prod_{1\leq s<t \leq N}\left\{
(a'_s-a'_t)^2-\frac{1}{4}(n_s-n_t)^2
\right\}
e^{\frac{8\pi^2i}{g_{\rm YM}^2A}\sum_{s=1}^N
\left({a'_s}^2
+\frac{n_s^2}{4}
\right)},
\ee
where $a'_s=a_s-\frac{1}{N}a,\;\;a=\sum_sa_s$ and the integral over $a$ has been performed.
Irrelevant constants and divergences are absorbed into a renormalized constant $C$.
In this limit, the poles in the integral measure have disappeared, then we have taken integral domains as whole space of integral variables $a'_s$.
By rescaling $a'_s$ by $y_s \equiv 2 a'_s$ and making 
an analytical continuation $g_{YM}^2 \rightarrow -i g_{YM}^2$,
we finally obtain
\begin{align}
\calZ_{N} =C'\sum_{\sum_sn_s=0}
\int_{\sum_sy_s=0} \prod_{s=1}^N dy_s
\prod_{1\leq s<t \leq N}\left\{
(y_s-y_t)^2-(n_s-n_t)^2
\right\}
e^{-\frac{2\pi^2}{g_{\rm YM}^2 A}\sum_{s=1}^N
\left(
y_s^2
+n_s^2
\right)},
\label{MP partition function}
\end{align}
where irrelevant constants are again absorbed into a constant $C'$.
${\cal Z}_N$ exactly agrees with the partition function of 2d $SU(N)$ YM on $S^2$ 
\cite{Migdal:1975zg,Witten:1992xu,Minahan:1993tp,Gross:1994mr,Blau:1993hj}\footnote{Note that
(3.7) in \cite{Minahan:1993tp} represents the partition function of $U(N)$ YM on $S^2$.
By applying  the procedure in \cite{Minahan:1993tp} to the partition function of  $SU(N)$ YM
in \cite{Blau:1993hj},
it is easy to see that the corresponding expression of the partition function of $SU(N)$ YM takes 
the form (\ref{MP partition function}).}.

The physical meaning of the integers $n_s$ can be understood from the following argument.
The localization theorem in the continuum $SU(N)$ YM on $S^2$ \cite{Witten:1992xu,Blau:1993hj} 
says that the path integral of the
partition function is localized at the solutions of the classical equation of motion
\begin{align}
D_\mu f^{\mu\nu}=0,
\label{equation of motion}
\end{align}
which are given by (\ref{monopole solution}).
Substituting the solution (\ref{monopole solution}) into the YM action (\ref{2d YM}) which gives
the equation of motion (\ref{equation of motion})
yields
\begin{align}
S_{2dYM}= \frac{\mu^4}{g_{\rm YM}^2}\int_{S^2} \frac{d\Omega_2}{\mu^2} \Tr\left(
\frac{1}{4}f^{\mu\nu}f_{\mu\nu}
\right) 
=\frac{2\pi^2}{g_{\rm YM}^2 A}\sum_{s=1}^N n_s^2.
\label{action with theta term}
\end{align}
This coincides with the exponent appearing in (\ref{MP partition function}).
Thus we can identify the fluctuations of the size of blocks $n_s$ with the monopole charges of the classical solution, which is consistent with the classical equivalence reviewed in section 2 and suggests 
that the localization works for the matrix model in a manner analogous to the case of 
the continuum field theory.

\section{Conclusion and discussion}

In this Letter, we study the ${\cal N}=1^*$ matrix model which is obtained by
dimensionally reducing CS theory on $S^3$.
We decompose the matrix integral into the sectors classified by the representation of 
$SU(2)$. We show that the $N$-blocks sectors reproduce 2d $SU(N)$ YM on $S^2$
in the large matrix size limit. 

We reproduced the partition function of 2d YM on $S^2$ from the ${\cal N}=1^*$ matrix model.
It is relevant to investigate whether the correlation functions of the physical observables 
in 2d YM on $S^2$ can be reproduced from the matrix model.
For instance, the vev of 
$\Tr_R e^{\Phi}$, where the trace is taken over a representation
$R$ of the matrix $\Phi$, is easily calculated in the matrix model .
This kind of the observables should be interpreted as a Wilson loop-like operator
in 2d YM. 

Our result suggests that the localization also works for the ${\cal N}=1^*$ matrix model
as for 2d YM theory. 
It has been discussed in \cite{Steinacker:2007iq} by using a different matrix model.
We need further investigation on the localization mechanism of  ${\cal N}=1^*$ matrix model
and relationship of our work to \cite{Steinacker:2007iq}.

We expect that CS theories on $S^3$ and the lens space $S^3/Z_q$ are obtained from the ${\cal N}=1^*$ matrix model
also at quantum level as ${\cal N}=4$ super-YM on $R\times S^3$ is obtained
from the plane wave matrix model \cite{Ishii:2008ib,Ishiki:2008te,Kitazawa:2008mx}. 
In this case, the operator $\Tr_R e^{\Phi}$ in the matrix model should correspond to
the Wilson loop operator in CS theory \cite{Ishii:2007sy}, and hopefully the knot invariant is derived from 
the matrix model.

The ${\cal N}=1^*$ matrix model is also interesting from the point of view of 4d super-YM theory, since the large $N$ limit of
the matrix model describes the effective superpotential of ${\cal N}=1^*$
theory which is mass deformed theory from ${\cal N}=4$ theory \cite{Dijkgraaf:2002fc,Dorey:2001qj}.
The different sectors of the $SU(2)$ representations that we have investigated should be related
to the different Higgs branches of the  ${\cal N}=1^*$ theory. The effective superpotential in the
different Higgs branches can be investigated by using the direct integration of the matrix model
partition function.

While we have extracted the $N$-block sectors `by hand' in the present Letter, we may
expect that the large $N$
limit with the large $N_0$ limit 
realizes 2d large $N$ YM on $S^2$ naturally, as the planar limit of ${\cal N}=4$ super-YM  
is realized in \cite{Ishii:2008ib}. 
The large $N$ limit seems relevant for the 
following reason.
It has been already pointed out that the $1/N$ expansion of 2d YM describes the genus expansion of
(non-critical) string theory \cite{Gross:1992tu}. One can 
deduce a world-sheet description of string theory from
the partition function of 2d YM. On the other hand, in this Letter, we have derived
the partition function of 2d YM  from the ${\cal N}=1^*$ matrix model in the large matrix size limit.
Matrix models are often regarded as regularization of
(non-critical) string theory, giving  world-sheet description in the large matrix size limit.
Our investigation strongly suggests the relationship between the matrix models in the large matrix 
size limit,  gauge theory
in the large $N$ limit
and string theory and the ${\cal N}=1^*$ matrix model is a good example
to understand the relationship (see also \cite{Dijkgraaf:2003xk,Dorey:2003pp}).
Further investigation of the ${\cal N}=1^*$ matrix model may
shed lights on nonperturbative definition of string theory.

\section*{Acknowledgements}
K.O. would like to thank participants of string theory meeting 2008 at RIKEN
for useful discussions and comments.
The work of G.I. and S.S. is supported in part by the JSPS Research Fellowship for Young
Scientists.
The work of K.O. and A.T. is supported in part by Grant-in-Aid for Scientific
Research (Nos.19740120 and 19540294) from the Ministry 
of Education, Culture, Sports, Science and Technology, respectively.


\appendix

\section{$S^3$ and $S^2$}
In this appendix, we summarize some useful facts about $S^3$ and $S^2$ 
(See also \cite{Ishii:2008tm,Ishii:2008ib}).
$S^3$ is viewed as the $SU(2)$ group manifold. We parameterize an element
of $SU(2)$ in terms of the Euler angles as
\begin{equation}
g=e^{-i\varphi \sigma_3/2}e^{-i\theta \sigma_2/2}e^{-i\psi \sigma_3/2},
\label{Euler angles}
\end{equation}
where $0\leq \theta\leq \pi$, $0\leq \varphi < 2\pi$, $0\leq \psi < 4\pi$.
The periodicity with respect to these angle variables is expressed as
\begin{align}
(\theta,\varphi,\psi)\sim (\theta,\varphi+2\pi,\psi+2\pi)\sim (\theta,\varphi,\psi+4\pi).
\end{align}
The isometry of $S^3$ is $SO(4)=SU(2)\times SU(2)$, and these two
$SU(2)$'s act on $g$ from left and right, respectively. 
We construct the
right-invariant 1-forms, 
\begin{equation}
dgg^{-1}=-i\mu E^i \sigma_i/2,
\label{right-invariant 1-form}
\end{equation}
where the radius of $S^3$ is given by $2/\mu$. They are explicitly
given by 
\begin{eqnarray}
&&E^1=\frac{1}{\mu}(-\sin \varphi d\theta + \sin\theta\cos\varphi d\psi),\nonumber\\
&&E^2=\frac{1}{\mu}(\cos \varphi d\theta + \sin\theta\sin\varphi
 d\psi),\nonumber\\
&&E^3=\frac{1}{\mu}(d\varphi + \cos\theta d\psi),
\end{eqnarray}
and satisfy the Maure-Cartan equation
\begin{equation}
dE^i-\frac{\mu}{2}\epsilon_{ijk}E^j\wedge E^k=0.\label{Maure-Cartan}
\end{equation}
The metric is constructed from $E^i$ as
\begin{equation}
ds^2=E^iE^i=\frac{1}{\mu^2}\left(
d\theta^2+\sin^2\theta d\varphi^2 +(d\psi+\cos\theta d\varphi)^2\right).
\label{metric of S^3}
\end{equation}
The Killing vector dual to $E^i$ is given by
\begin{equation}
{\cal{L}}_i=-\frac{i}{\mu}E^M_i\partial_M,
\label{definition of Killing vector}
\end{equation}
where $M=\theta,\varphi,\psi$ and $E^M_i$ are inverse of $E^i_M$. The explicit form of the Killing
vector is 
\begin{eqnarray}
&&{\cal{L}}_1=-i\left(-\sin\varphi\partial_{\theta}-\cot\theta\cos\varphi\partial_{\varphi}+\frac{\cos\varphi}{\sin\theta}\partial_{\psi}\right),\nonumber\\
&&{\cal{L}}_2=-i\left(\cos\varphi\partial_{\theta}-\cot\theta\sin\varphi\partial_{\varphi}+\frac{\sin\varphi}{\sin\theta}\partial_{\psi}\right),\nonumber\\
&&{\cal{L}}_3=-i\partial_{\varphi}.\label{Killing vector}
\end{eqnarray}
Because of the Maure-Cartan equation (\ref{Maure-Cartan}), the Killing vector satisfies 
the $SU(2)$algebra $[{\cal{L}}_i,{\cal{L}}_j]=i\epsilon_{ijk}{\cal{L}}_k$.

One can also regard $S^3$ as a $U(1)$ bundle over $S^2=SU(2)/U(1)$.
$S^2$ is parametrized by $\theta$ and $\varphi$ and covered with two local patches:
the patch I defined by $0\leq\theta <\pi$ and the patch II defined by $0<\theta\leq\pi$.
In the following expressions, the upper sign is taken in the patch I while the lower sign in the patch II.
The element of $SU(2)$ in (\ref{Euler angles}) is decomposed as 
\begin{align}
g=L\cdot h
\end{align}
with
\begin{align}
&L=e^{-i\varphi \sigma_3/2}e^{-i\theta \sigma_2/2}e^{\pm i\varphi \sigma_3/2}, \nonumber\\ 
&h=e^{-i(\psi\pm\varphi) \sigma_3/2}. 
\end{align}
$L$ represents an element of $S^2$, while $h$
represents the fiber $U(1)$.
The fiber direction is parametrized by $y=\psi\pm\varphi$.
Note that $L$ has no $\varphi$-dependence for $\theta=0,\pi$.
The zweibein of $S^2$ is given by the $i=1,2$ components of the left-invariant 1-form, 
$-iL^{-1}dL=\mu e^i\sigma_i/2$.
It takes the form
\begin{align}
&e^1=\frac{1}{\mu}(\pm\sin\varphi d\theta+\sin\theta\cos\varphi d\varphi), \nonumber\\
&e^2=\frac{1}{\mu}(-\cos\varphi d\theta \pm \sin\theta\sin\varphi d\varphi).
\end{align}
This zweibein gives the standard metric of $S^2$ with the radius $1/\mu$:
\begin{align}
ds^2=e^ie^i=\frac{1}{\mu^2}(d\theta^2+\sin^2\theta d\varphi^2).
\label{metric of S^2}
\end{align}
Making a replacement $\partial_y \rightarrow -iq$ in (\ref{Killing vector})
leads to the angular momentum operator in the presence of a monopole with
magnetic charge $q$ at the origin \cite{Wu:1976ge}:
\begin{eqnarray}
&&L_1^{(q)}=i(\sin\varphi\partial_{\theta}+
\cot\theta\cos\varphi\partial_{\varphi})-
q\frac{1\mp \cos\theta}{\sin\theta}\cos\varphi, \nonumber\\
&&L_2^{(q)}=i(-\cos\varphi\partial_{\theta}+
\cot\theta\sin\varphi\partial_{\varphi})-
q\frac{1\mp \cos\theta}{\sin\theta}\sin\varphi, \nonumber\\
&&L_3^{(q)}=-i\partial_{\varphi}\mp q ,
\label{monopole angular momentum}
\end{eqnarray}
where $q$ is quantized as 
$q=0, \pm \frac{1}{2}, \pm 1, \pm \frac{3}{2},\cdots$,
because $y$ is a periodic variable with the period $4\pi$.
These operators act on the local sections on $S^2$ and satisfy the $SU(2)$
algebra $[L_i^{(q)},L_j^{(q)}]=i\epsilon_{ijk}L_k^{(q)} $, 
Note that when $q=0$, these operators are reduced to the ordinary
angular momentum operators (\ref{monopole angular momentum}) on $S^2$ (or $R^3$),
which generate the isometry group of $S^2$, $SU(2)$.
The $SU(2)$ acting on $g$ from left survives as the isometry of $S^2$.

\end{document}